\documentclass[a4paper,12pt]{article}
\usepackage{graphicx}
\usepackage{epsfig}
  
\newcommand{\ds}{\displaystyle }
\newcommand{\R}{{\sf R\hspace*{-0.9ex}%
\rule{0.15ex}{1.5ex}\hspace*{0.9ex}}}
\newcommand{\Z}{{\sf Z\hspace*{-0.9ex}%
\rule{0.15ex}{1.5ex}\hspace*{0.9ex}}}
\newcommand{\N}{{\sf N\hspace*{-0.9ex}%
\rule{0.15ex}{1.5ex}\hspace*{0.9ex}}}

\title{An approach to anomalous diffusion in the $n$-dimensional space generated by a self-similar Laplacian}

\author{ {\sl Thomas M. Michelitsch$^{1}$\footnote{Corresponding author, e-mail~: michel@lmm.jussieu.fr },
G\'{e}rard A. Maugin$^{1}$} \\ \\ {\sl Andrzej F. Nowakowski$^2$, Franck C. G. A. Nicolleau$^2$}
\\ \\ {\sl Mujibur Rahman$^3$} \\ \\
$^1$ Universit\'{e} Pierre et Marie Curie, Paris 6\\
Institut Jean le Rond d'Alembert \\ CNRS UMR 7190 \\ France
\\ \\
$^2$ Sheffield Fluid Mechanics Group\\
Department of Mechanical Engineering\\
University of Sheffield\\
United Kingdom
\\ \\
$^3$ GE Energy, Greenville, SC 29615\\
USA
\\ \\ {\it submitted manuscript}\\ {\small \it \jobname .tex }
}
\begin{document}

\maketitle

\newpage

\textbf{Abstract}
We analyze a quasi-continuous linear chain with self-similar distribution of harmonic interparticle springs as introduced for one dimension in \cite{michel}. We define a continuum limit for one dimension and generalize it to $n=1,2,3,..$ dimensions of the physical space. Application of Hamilton's (variational) principle defines then a self-similar and as consequence non-local Laplacian operator for the $n$-dimensional space where we proof its ellipticity and its accordance (up to a strictly positive prefactor) with the {\it fractional Laplacian} $-(-\Delta)^\frac{\alpha}{2}$. By employing this Laplacian we establish a Fokker Planck diffusion equation: We show that this Laplacian generates spatially isotropic L\'evi stable distributions which correspond to L\'evi flights in $n$-dimensions.
In the limit of large scaled times $\sim t/r^{\alpha} >>1$ the obtained distributions exhibit an algebraic decay $\sim t^{-\frac{n}{\alpha}} \rightarrow 0$ independent from the initial distribution and spacepoint. This universal scaling depends only on the ratio $n/\alpha$ of the dimension $n$ of the physical space and the L\'evi parameter $\alpha$.
\newline\newline
{\sl Keywords. Fractional Laplacian, Fokker Planck equation, anomalous diffusion, L\'evi flights, L\'evi (stable) distributions, self-similarity, scaling laws, fractional operator, non-locality}

\section{Introduction}

Self-similarity (scaling-invariance) can be found in many problems of physics. 
Hence there is a need for understanding the dynamics that leads to fractal and self-similarity properties in domains of the physics as varied as flow turbulence and complex materials.
In `traditional modelling' the continuum is assumed to have a characteristic length scale which determines the wavelengths where wave fields interact with the
microstructure. However, there are numerous materials in nature which are constituted by a scale hierarchy of
recurrent microstructure which can be conceived in a good approximation as self-similar. This is true for solids and porous media but also in fluid mechanics, or multicomponent and multiphase flows.
As a consequence, there are many areas of modelling that could benefit from a better understanding of the mechanisms underlying the dynamics of objects exhibiting self-similarity properties.
In fluid mechanics, synthetic turbulence models are one of many examples of tempering with the input spectrum in order to understand better the physics underlying Lagrangian diffusion
\cite{Nicolleau-et-al-erc-2011} and such spectra can be traced to the fractal distributions of velocity accumulation points in the synthetic flow.
Fractal approaches are developing rapidly in fluid mechanics. Such approaches consist in either experimentally or numerically interfering with the flow, forcing it through self-similar objects (or through numerical forcing) \cite{Seoud-Vassilicos-2007,Hurst-Vassilicos-2007,Nicolleau-et-al-JOT-2011}.

To the best of our knowledge there is no generally accepted {\it continuum field approach} that is able to take into account self-similarity (scaling invariance). Therefore, simple continuum models accounting for fractal and self-similar mechanisms are highly desirable. Strictly speaking, so far there has not been generally accepted procedure describing how key operators such as the Laplacian have to be modified when self-similarity of particle interactions comes into play.
The recent paper \cite{michel} develops the first steps for such a procedure. However, limited to the one-dimensional case. In that paper we established a ``self-similar Laplacian operator'' from the elastic energy of a quasi-continuous linear chain with self-similar particle interactions.
This was followed by the derivation of a continuum limit for that Laplacian \cite{michel2}. In the present paper we briefly revisit this model and show that this self-similar Laplacian as well as its generalization to $n$ dimensions can be rigorously obtained by application of Hamilton's variational principle.
Furthermore, we generalize the Laplacian to the physical space in $n$ dimension where spatial isotropy of the Laplacian is maintained. The diffusion processes generated by a Fokker Planck equation is analysed using the self-similar Laplacian. It is shown that
independently of $n$, the physical dimension, the underlying anomalous diffusion processes are L\'evi flights. In this way the present model could be an essential feature and necessary ingredient for flow turbulence.
\\[2ex]
It has been recognized that complex processes exist where the role of fluctuations is by far underestimated and which cannot be described by Gaussian statistics. These random particle motions have much more erratic characteristics as the ``benign'' Brownian motions which can be described by Gaussian statistics and normal distributions.
This is true for instance for the stock market, where Gaussian statistics fails largely by underestimating fluctuations (risks!). One of the first researchers to recognize this was indeed Mandelbrot \cite{mandel,mandel2}.
Among the stochastic motions that are characterized by infinite mean fluctuations are those with L\'evi-distributed scale free jump distributions. Such motions, known in the literature as L\'evi flights are the source of anomalous diffusion \cite{levi}. L\'evi flights are widely found in nature for instance in the dynamics of bumblebees \cite{klages}. There is a huge number of models of L\'evi flights by fractional kinetic equations of diffusion and a vast literature devoted to this subject. An excellent overview is provided in \cite{metzler} (see also the references therein). However, the subject of the present paper is not the investigation of the properties of L\'evi flights themselves which are well studied, e.g. \cite{gloeckle,sun,metzler}. In most of these models a diffusion problem is ad hoc {\it defined} such that L\'evi stable distributions solve the fractional problem. However, a physical interpretation or justification of these fractional operators is in many cases
  not given \cite{gloeckle,sun}. The goal of the present paper is to establish a simple and physically well based model where the Laplacian used in the Fokker Planck diffusion equation is rigorously deduced from Hamilton's principle.

This paper is organized as follows: We deduce a self-similar Laplacian as a result of a continuum limit of the linear chain with a self-similar distribution of interparticle springs in one dimension. This continuum limit results in a Laplacian in the form of a self-adjoint combination of fractional operators.
Then we generalize this Laplacian to the $n$-dimensional physical space ($n=1,2,3$). The generalization to higher dimensions than one is performed such that the Laplacian maintains spatially isotropic symmetry.
We deduce the dispersion relation and the density of normal modes (oscillator density) both obeying characteristic scaling laws. By employing this Laplacian we formulate a diffusion equation (Fokker Planck equation) which generates L\'evi stable distributions. It does not matter for the model whether we conceive this distribution as particle density function or probability density function. In the first case we consider an ensemble of diffusing particles and in the other case the probability distribution of one propagating particle. For the latter case the stochastic particle motion which describes our model are L\'evi flights. Despite we refer more to the ``diffusion picture'', the interpretation of our model can be fully transferred into the ``stochastic picture'' of a single randomly walking L\'evi flyer particle.

We derive the time evolution operator in its space-time representation. We focus especially on the asymptotic behavior at large (scaled) times
which is found to be a spatially uniform algebraic decay approaching zero density. We demonstrate that this characteristic spatially homogeneous behavior of the asymptotic density indicates the approach of the system to a quasi-equal distribution approaching
maximum (infinite) entropy due to the complete uncertainty of the location of a particle independent from the initial distribution.

To deduce the self-similar Laplacian operator we consider a ``material system'' with a spatially {\it homogeneous} constant mass distribution and spatially harmonic {\it self-similar interparticle interactions}. As demonstrated in \cite{michel} such a material system exhibits a regime with a fractal dispersion relation that appears in a Weierstrass-Mandelbrot function which is exactly self-similar. Interparticle interactions which are exactly self-similar in the below defined sense require an infinite physical space.
In the continuum limit of this model
the fractal properties of the dispersion relation disappear and the dispersion relation assumes the form of a smooth power function \cite{michel,michel2a,michel2}. The continuum limit of this model opens a general way to deduce continuous field theories in media with self-similar interparticle interactions leading to self-similar constitutive laws. In a recent paper \cite{michel2a} this theory is elaborated for one dimension. In the present paper we
generalize this model to $n=1,2,3,..$ dimensions of the physical space. For the sake of simplicity we refer to the medium with self-similar interactions just as ``self-similar medium'' or ``self-similar continuum''.
We emphasize that the self-similarity of interparticle interactions is the inevitable source of non-locality in the deduced Laplacian operator. In contrast to this, a linear chain which contains only harmonic next neighbor interparticle interactions yields in the continuum limit the traditional spatially local Laplacian.
A discrete material system leading to a {\it fractal dispersion relations} due to non-local interparticle interactions has probably first been considered by Tarasov \cite{tarasov1} (see also the numerous references therein).

\section{Preliminary remarks}

It is an interesting question how physical phenomena change when the interparticle interactions become self-similar. For instance when we consider the Poisson equation

\begin{equation}
 \label{laplaceeq} \Delta_{\mathrm{selfsim}} \, u({\bf r}) = \chi({\bf r})
\end{equation}
with a {\it self-similar Laplacian operator} $\Delta_{\mathrm{selfsim}}$ which is to be specified.
To this end we should elaborate first of all the notion of ``self-similarity'' employed in this paper.
The notion of ``{\it self-similarity}'' which is employed in this paper as well as in \cite{michel,michel2a,michel2} corresponds to the notion "{\it self-similarity at a point}" which is commonly used in the mathematical literature \cite{peitgen}. Generally, an object is exactly self-similar in the strict sense if it can be decomposed into parts which are exact rescaled copies of the entire object. In contrast is the notion "self-similarity at a point" where ``point'' means here a fixed-point of the scaling operation: An object which is {\it self-similar at a point contains a single part} which is a re-scaled copy of the entire object and so forth over an infinity of scales \cite{peitgen}.

According to this notion of self-similarity we call a function $\Lambda(h)$ self-similar with respect to $h$, i.e. self-similar at point $h=0$, when the relation

\begin{equation}
 \label{selfunction} \Lambda(Nh)=N^{\delta}\Lambda(h)
\end{equation}
is fulfilled for a prescribed scaling factor $N>1$ and for {\it any $h>0$}. We assume real valued scaling exponents $\delta \in \R$.
If a function $\Lambda(h)$ fulfills (\ref{selfunction}), it follows that (\ref{selfunction}) remains true if we replace $N \rightarrow N^s$ with ($s\in \Z_0$ denotes positive and negative integers including zero). In other words: if $\Lambda(h)$ is self-similar with respect to $h$ in the sense of relation (\ref{selfunction}), then there exists a $N>1$ such that {\it the discrete set} of rescaling operations $h'=hN'$ with $N'=N^s$ with {\it only positive and negative integers $s\in \Z_0$ including the zero}, satisfy the self-similarity condition (\ref{selfunction}), namely $\Lambda(hN')=N'^{\delta}\Lambda(h)$\footnote{In the continuum limit to be discussed below, this set $\{hN^s\}$ becomes a continuous one.}.
We observe further by putting $h=N^{p+\chi}$ with $p\in \Z_0$ and $0\leq \chi <1$ denoting the non-integer part, and by using the property of self-similarity that ($N^{p}=hN^{-\chi}$)

\begin{equation}
\label{hoelder}
\Lambda(h=N^{p}N^{\chi})=N^{p\delta}\Lambda(N^{\chi})=h^{\delta}N^{-\delta\chi}\Lambda(N^{\chi})
\end{equation}
From this relation we observe that all values of $\Lambda$ are uniquely determined by its values
within $1\leq N^{\chi} <N$ (as $0\leq \chi <1$) \cite{michel-wittenberg}. Further we see that $\Lambda(h)$ scales as $h^{\delta}$, especially when $h\rightarrow 0$. However, in general, a unique limit $h\rightarrow 0$ of $h^{-\delta}\Lambda(h)=N^{-\delta\chi}\Lambda(N^{\chi})$ does not exist because of its dependence on $\chi$. Let us assume that a constant $C>0$ exists such that $0<|N^{-\delta\chi}\Lambda(N^{\chi})| \leq C$ $\forall h> 0$ then function $\Lambda(h)$ fulfills the inequality

\begin{equation}
\label{hoelderineq}
0< |\Lambda(h)| \leq C h^{\delta}
\end{equation}
For $0<\delta\leq 1$ relation (\ref{hoelderineq}) is the {\it Hoelder condition}, e.g. \cite{han}. The function $\Lambda(h)$ is then a Hoelderian function (Hoelder continuous function) being continuous but non-differentiable for $0<\delta<1$ at $h=0$. Hoelderian functions include a wide range of fractal and erratic functions \cite{han}. For $\delta \geq 1$ the function $\Lambda(h)$ is differentiable at $h=0$. Self-similar functions can hence be fractal or non-fractal functions.

Generally a self-similar function which fulfills (\ref{selfunction}) for a prescribed $N$ can be written in the form
\begin{equation}
\label{selfsimfunc}
\Lambda(h)=\sum_{s=-\infty}^{\infty}N^{-\delta s}f(N^sh)
\end{equation}
which converges for sufficiently good functions $f$ \cite{michel}.
Without any loss of generality we can restrict ourselves to $N>1$. The simplest self-similar functions of this type are power-functions $h^{\delta}$. They constitute also the continuum limit of (\ref{selfsimfunc}) (eq. (\ref{selfu}) below).

\section{The self-similar elastic continuum}
\label{1D}
\subsection{One-dimensional case}

In this section we evoke a self-similar Laplacian from a simple linear chain model which was introduced in \cite{michel}
and its continuum limit we introduced in recent papers \cite{michel2a,michel2} for the one-dimensional infinite space.
We consider this quasi-continuous chain with self-similar harmonic springs with
the Hamiltonian functional \cite{michel}

\begin{equation}
\label{Hamiltonian}
H = \frac{1}{2}\int_{-\infty}^{\infty}\left(\dot{u}^2(x,t)+{\cal V}(x,t,h)\right){\rm d}x
\end{equation}
where $x$ denotes the space- and $t$ the time coordinates.
Each spacepoint $x$ denotes also a mass point and the mass density
of the system we consider is constant (spatially homogeneous) and we put the mass density equal to $1$.
$\frac{1}{2}{\cal V}(x,t,h)$ indicates the elastic energy density with

\begin{equation}
\label{elastic}
\ds {\cal V}(x,t,h)= \frac{1}{2}\sum_{s=-\infty}^{\infty}N^{-\delta s}\left[\left\{u(x,t)-u(x+hN^s,t)\right\}^2+
\left\{u(x,t)-u(x-hN^s,t)\right\}^2\right]
\end{equation}
which converges in the range $0<\delta<2$ and where we assume $h>0$ and $N$ being a prescribed scaling factor. Without loss of generality we can restrict ourselves to $N>1$ ($N\in \R$).
The additional factor ${1}/{2}$ in the elastic
energy (\ref{elastic}) compensates double counting of the springs when integrating in (\ref{Hamiltonian}).
It is important to note that the elastic energy density does not have any characteristic interaction length scale. The variable $h$ characterizes self-similarity of the elastic energy density, but does not have the physical meaning of a characteristic length. Unlike in the case of a linear chain with only next neighbor interactions, the limit $h\rightarrow 0$ would not localize the
interparticle interactions.
In (\ref{Hamiltonian}) and (\ref{elastic}) $u$ and $\dot{u}=\frac{\partial }{\partial t}u$ stand for the displacement field and the velocity field, respectively. (\ref{elastic}) has the property of being self-similar with respect to $h$ at point $h=0$, namely

\begin{equation}
\label{selfsimh0}
{\cal V}(x,t,Nh)=N^{\delta}{\cal V}(x,t,h)
\end{equation}

As a starting point for the approach to be developed we evoke the continuum limit of (\ref{elastic})
and the resulting equation of motion. For $0<\delta \leq 1$ (\ref{elastic}) is a Hoelder continuous function being for $0<\delta<1$ non-differentiable at $h=0$. If we prescribe in (\ref{elastic}) a periodic field $u(x)$ then the elastic energy density is in $0<\delta \leq 1$ a {\it fractal} function (Weierstrass-Mandelbrot fractal function).
We define the continuum limit as $N=1+\zeta$ ($0<\zeta<<1$) so $\tau=hN^s$ becomes a continuous variable and we can
write a self-similar function $\Lambda(h)$ which fulfills a self-similarity condition (\ref{selfsimh0}) asymptotically as \cite{michel}

\begin{equation}
\label{selfu}
\Lambda(h)=\sum_{s=-\infty}^{\infty}N^{-\delta s}f(N^sh) \sim \frac{h^\delta}{\zeta}\int_0^{\infty}\frac{f(\tau)}{\tau^{\delta +1}}{\rm d}\tau
\end{equation}
having the form of a power function $\Lambda(h)=const\,h^{\delta}$.
Both the discrete as well as the continuous representation of (\ref{selfu}) converge for sufficiently good functions (see details in \cite{michel}).
From (\ref{selfu}) follows that in that continuum limit we can write (\ref{elastic}) as a functional of the displacement field $u(x,t)$ in the form

\begin{equation}
\label{elastfunc}
\ds {\cal V}(x,t,h)= \frac{h^\delta}{2\zeta}\int_0^{\infty}\frac{\left\{(u(x,t)-u(x+\tau,t))^2+
(u(x,t)-u(x-\tau,t))^2\right\}}
{\tau^{\delta +1}}{\rm d}\tau
\end{equation}
which exists as in the discrete case (\ref{elastic}) in the band $0<\delta<2$. Application of Hamilton's principle leads then to the definition of the {\it Laplacian} of our system which is then determined by the
functional derivative of the elastic energy $V$ with respect to the field $u$, namely \cite{goldstein}

\begin{equation}
\label{varprinc}
\Delta_{(\delta,h,\zeta)}u =: -\frac{\delta V}{\delta u}, \hspace{1cm} V=\frac{1}{2}\int {\cal V}{\rm d}x
\end{equation}
The equation of motion (self-similar wave equation) has then the form \cite{michel}

\begin{equation}
\label{fractlapself} \frac{\partial^2 }{\partial t^2}u(x,t) = \Delta_{(\delta,h,\zeta)}u(x,t)
\end{equation}

 (\ref{varprinc}) together with (\ref{elastfunc}) defines the {\it self-similar Laplacian} in its continuous representation of the one-dimensional medium. Relation
 (\ref{varprinc}) defines the Laplacian for both cases, in the the discrete case of (\ref{elastic}) as well as for the continuous case (\ref{elastfunc}).
 In both cases the Laplacian is due to its construction self-similar with respect to $h$.
 The self-similar Laplacian is necessarily a non-local and {\it self-adjoint} negative semi-definite\footnote{Uniform translations are eigenmodes to eigenvalue zero.}, spatially isotropic operator: In the continuous case it is obtained as

\begin{equation}
\label{laplace}
\ds \Delta_{(\delta,h,\zeta)}u(x)= \frac{h^{\delta}}{\zeta}\int_0^{\infty}\frac{(u(x-\tau)+u(x+\tau)-2u(x))}{\tau^{1+\delta}}\,{\rm
d}\tau
\end{equation}
existing for $0<\delta<2$. It might be sometimes convenient to rewrite (\ref{laplace}) in the equivalent form
\begin{equation}
\label{laplaceb}
\Delta_{(\delta,h,\zeta)}u(x)=\frac{h^{\delta}}{\zeta\delta}\frac{d }{d x}\int_0^{\infty}\frac{(u(x+\tau)-u(x-\tau))}{\tau^{\delta}}\,{\rm
d}\tau \,,\hspace{2cm} 0<\delta<2
\end{equation}
where the range of existence of these relations is $0<\delta<2$.
In the entire analysis of this paper we put for any complex number $z=|z|e^{i\varphi}$ the {\it principal value} $-\pi < \varphi=Arg(z) \leq \pi$ for its argument $\varphi$.
In the further analysis we will need the $\Gamma$-function (faculty-function) $\Gamma(z)$ which is defined by \cite{abramowitz}

\begin{equation}
\label{gammafu}
\Gamma(\alpha+1)=:\alpha!=\int_0^{\infty}e^{-\tau}\tau^{\alpha}{\rm d}\tau ,\,\,\, Re(\alpha) > -1
\end{equation}
The condition $Re(\alpha)>-1$ is required for integral (\ref{gammafu}) to exist.
$Re(Z)$ denotes the real- and $Im(Z)$ the imaginary part of a complex number $Z$.
Using (\ref{laplaceb}) the equation of motion (\ref{fractlapself}) takes the form

\begin{equation}
 \label{divstress}
\frac{\partial^2 }{\partial t^2}u(x,t) =\frac{\partial}{\partial x} \sigma(x,t)
\end{equation}
where $\frac{\partial}{\partial x}$ indicates the traditional partial derivative with respect to $x$ and $\sigma(x,t)$ denote the stress having the form \cite{michel2}

\begin{equation}
 \label{stresstensor}
\sigma(x)= \frac{h^{\delta}}{\zeta\delta}\int_0^{\infty} \frac{(u(x+\tau)-u(x-\tau))}{\tau^{\delta}}\,{\rm
d}\tau \,,\hspace{1cm} 0<\delta<2
\end{equation}
where the integration constant turns out to be zero when integrating the right hand side of (\ref{laplaceb}) with respect to $x$.
For further comparison it will be convenient to represent (\ref{stresstensor}) in the equivalent form
\begin{equation}
 \label{stresstensorb}
\sigma(x)= \frac{h^{\delta}}{2\zeta\delta}\int_{-\infty}^{\infty} \tau\frac{(u(x+\tau)-u(x-\tau))}{|\tau|^{\delta+1}}\,{\rm
d}\tau \,,\hspace{1cm} 0<\delta<2
\end{equation}
where $sgn(\tau)=\frac{\tau}{|\tau|}$ maintains the integrand to be an even function with respect to $\tau$.

From this follows that self-similarity of the interparticle interactions leads inevitably to non-local field theories. In the elastic framework the material (moduli) functions are convolution kernels. In contrast to the ``classical'' non-local elasticity theory as
outlined by Eringern \cite{eringen} the self-similar case is characterized by long-range power law kernels decaying critically slowly \cite{michel2a,michel2}.

\section{The self-similar Laplacian of the $n$-dimensional space}
\label{ndimlaplace}

We introduce a generalization of the self-similar Laplacian (\ref{laplace}) to the $n$-dimensional space where $n=1,2,3$. We define this
Laplacian by its action on a scalar field variable $u({\bf x})$. Then we can
generalize the one-dimensional case (\ref{laplace}) to $n=1,2,3$ dimensions as\footnote{The multiplyer $h^{\delta}$ has been changed into $h^{\delta-(n-1)}$ to keep the dimension of the self-similar Laplacian independent on $n$.}

\begin{equation}
 \label{laplaceddim}
\ds \Delta_{(n,\delta)}u({\bf x})= \frac{h^{\delta-(n-1)}}{2\zeta}\int \frac{\left(u({\bf x}+{\bf r})+u({\bf x}-{\bf r}) -2u({\bf x})\right)} {r^{\delta+1}}{\rm d}^n{\bf r} ,\,\hspace{1cm} 0 <\delta-(n-1) < 2
\end{equation}

This integral is performed over the entire $n$-dimensional physical space $\R^n$. The prefactor $2^{-1}$ has been chosen due to the fact that in (\ref{laplace}) a prefactor $2^{-1}$ has to be added to the integrand if we integrate over the entire physical space $\R^1$. (\ref{laplaceddim})
exists for sufficiently smooth fields $u({\bf x})$ converging in the interval $n-1 < \delta <n+1$.
In some cases also the equivalent representation

\begin{equation}
 \label{laplaceddimb}
\ds \Delta_{(n,\delta)}u({\bf x})= \frac{h^{\delta-(n-1)}}{\zeta}\int \frac{\left\{u({\bf r})-u({\bf x})\right\}} {|{\bf r}-{\bf x}|^{\delta+1}}{\rm d}^n{\bf r} ,\,\hspace{1cm} 0 <\delta-(n-1) < 2
\end{equation}
might be convenient. In the appendix (\ref{appAdeltan}) we demonstrate that our self-similar Laplacian (\ref{laplaceddim}),
(\ref{laplaceddimb}) is up to a strictly positive prefactor coinciding with the {\it fractional Laplacian} $-(-\Delta)^{\frac{\alpha}{2}}$  (with $\alpha=\delta-(n-1)$ and $0<\alpha<2$) known from the literature, e.g. \cite{vazquez} and the references therein.

For the analysis the following representation of (\ref{laplaceddim}) in terms of a divergence of a vector field will be useful
\begin{equation}
 \label{gausslaplace}
\ds \Delta_{(n,\delta)}u({\bf x}) = \nabla_{\bf x}\cdot{\bf D}({\bf x})
\end{equation}
where $(\nabla_{{\bf x}})_j=\frac{\partial}{\partial x_j}$ denotes the (traditional) gradient operator.
The vector field ${\bf D}$ is determined (up to an unimportant rotational gauge vector field ${\bf b}$ with $\nabla\cdot{\bf b}=0$) by
\begin{equation}
 \label{vecfield}
{\bf D}{(\bf x})= \frac{1}{(\delta-(n-1))}\frac{h^{\delta-(n-1)}}{2\zeta}\int
\frac{{\bf r}}{r^{\delta+1}} \left\{u({\bf x}+{\bf r})-u({\bf x}-{\bf r})\right\}\,{\rm d}^n{\bf r} ,\,\hspace{1cm} 0 <\delta-(n-1) < 2
\end{equation}
which recovers for $n=1$ expression (\ref{stresstensorb}).
This vector field exists in the same interval $n-1 < \delta <n+1$ as (\ref{laplaceddim}). The deduction of
(\ref{vecfield}) is performed in the appendix \ref{gausstheo} by using the Gaussian theorem.
Further useful equivalent representations of (\ref{vecfield}) are given in the appendix.
We note that $u$ is a scalar field and the {\it integrand} of (\ref{vecfield}) is an even function with respect to integration variable ${\bf r}$. For $n=1$ (\ref{vecfield}) corresponds to the stresses (\ref{stresstensorb}).
If we conceive equation (\ref{laplaceeq}) as a Poisson equation in an electrodynamic context then the vector field
(\ref{vecfield}) can be conceived as the ``dielectric displacement field''. Then (\ref{laplaceeq}) with (\ref{gausslaplace})
defines the self-similar Gauss-law (charge conservation).
Starting from this we can set up
a theory of self-similar fields and set up ``self-similar'' Maxwell equations. In the appendix \ref{gausstheo} we have deduced a useful scalar ''potential'' (eq. (\ref{poPhi})). However, we will elaborate the subject of a ``self-similar'' electrodynamic field theory in a sequel paper.

\subsection{Dispersion relation and density of normal modes}
\label{disprel}
In view of the translational symmetry of the Laplacian, we observe that plane-waves
$\phi_k({\bf r})= e^{i{\bf k}\cdot{\bf r}}$ are eigenfunctions of the Laplacian (\ref{laplaceddim}) where its negative eigenvalues constitute the dispersion relation $\omega^2(k)$ which is obtained by the relation

\begin{equation}
 \label{disporel}
\Delta_{(n,\delta)}\phi_k({\bf r})=-\omega_{n,\delta}^2(k)\phi_k({\bf r})
\end{equation}

\begin{equation}
 \label{dispo}
\omega_{n,\delta}^2(k)= -\frac{h^{\delta-(n-1)}}{2\zeta}\int\frac{\left(e^{i{\bf k}\cdot {\bf r}}+e^{-i{\bf k}\cdot {\bf r}} -2\right)} {r^{\delta+1}}{\rm d}^n{\bf r}
\end{equation}

We observe that the dispersion relation fulfills the scaling property

\begin{equation}
 \label{disperela}
\omega_{n,\delta}^2(k)=A_{n,\delta}k^{\delta-n+1} \,,\hspace{1cm} 0<\delta-n+1<2
\end{equation}
depending only on $k=|{\bf k}|$ reflecting isotropy of the Laplacian and $A_{n,\delta}=\omega_{n,\delta}^2(k=1)$ is defined by (\ref{dispo}). We observe from (\ref{dispo}) that
the coefficient $A_{n,\delta}>0$ is strictly positive indicating ``elastic stability''.
A further evaluation of the coefficient $A_{n,\delta}$ is given in the appendix \ref{appAdeltan}.
The following observation is noteworthy: {\it In dispersion relation (\ref{disperela}) appears always
a positive exponent $0<\delta-n+1<2$ being within the interval $(0,2)$ whatever the
dimension $n$ of the physical space}.
From this follows the important property
\begin{equation}
 \omega_{n,\delta}^2(k\rightarrow 0)=0
\end{equation}
for any dimension $n$ as a necessary consequence of the translational invariance of the self-similar Laplacian. Translational invariance requires that the $k=0$ mode (uniform translation of the entire ``material system'') must give a zero contribution to the elastic energy and hence corresponds to eigenvalue zero.

It is now straight-forward to obtain the {\it density of normal modes} which we denote by ${\cal D}(\omega)$.
This quantity is defined such that ${\cal D}(\omega){\rm d\omega}$
counts the number of normal modes (per $n$-dimensional unit-volume) with frequencies in the interval $[\omega,\omega +{\rm d}\omega]$.
We obtain this quantity from the dispersion relation by (e.g. \cite{michel} or any textbook of theoretical physics)

\begin{equation}
\label{oscilden}
{\cal D}(\omega){\rm d}\omega = \frac{1}{(2\pi)^n}O_n(1)k^{n-1}{\rm d}k
\end{equation}
where we put $\alpha=\delta-(n-1)$
where $k=k(\omega)=\frac{\omega^{\frac{2}{\alpha}}}{A^{\frac{1}{\alpha}}}$ is the inverse dispersion relation and $O_n(1)=\frac{2\pi^{\frac{n}{2}}}{\Gamma(\frac{n}{2})}$ denotes the surface of the unit-sphere. Then
(\ref{oscilden}) yields the scaling law

\begin{equation}
\label{oscdensity}
\ds {\cal D}(\omega) =\frac{2^{2-n}}{\pi^{\frac{n}{2}}\Gamma(\frac{n}{2})\alpha A_{n,\delta}^{\frac{n}{\alpha}}}\,\, \omega^{\frac{2n}{\alpha}-1}
\end{equation}
where the prefactor is always positive and the exponent $\frac{2n}{\alpha}-1 > n-1 \geq 0$ ($0<\alpha<2$) whatever the dimension $n$ of the physical space. Hence we have for $n=1$ an exponent $\frac{2}{\delta}-1>0$ ($0<\delta<2$); for $n=2$ an exponent
 $\frac{4}{\delta-1}-1>1$ ($1<\delta<3$); and for $n=3$ an exponent $\frac{6}{\delta-2}-1> 2$ ($2<\delta<4$).

The above power law (\ref{oscdensity}) confirms our conjecture raised in \cite{michel}, namely
that the property

\begin{equation}
 \label{propuniversel}
{\cal D}(\omega\rightarrow 0) = 0
\end{equation}
holds generally in self-similar media where the density of normal modes obeys a power law with positive exponent for any dimension $n$ of the physical space.
For $n=1$ the expression obtained in \cite{michel} is recovered by (\ref{propuniversel}) with the exponent
being $\frac{2}{\delta}-1 >0$ ($0<\delta<2$) and yields

\begin{equation}
\label{D1}
{\cal D}_{1}(\omega)=\frac{2}{\pi\delta A_{1,\delta}^{\frac{1}{\delta}}}\,\,\omega^{\frac{2}{\delta}-1}
\end{equation}

In contrast to the self-similar Laplacian of our model the traditional Laplacian would give a dispersion relation
$\omega^2(k)=k^2$ and correspondingly the density of normal modes would be in the $n$-dimensional space
\begin{equation}
 \label{traditional}
{\cal D}_{traditional}(\omega)= \frac{O_n(1)}{(2\pi)^n}\,\omega^{n-1} =\frac{2^{1-n}}{\pi^{\frac{n}{2}}\Gamma(\frac{n}{2})}\omega^{n-1}
\end{equation}
scaling as $\sim \omega^{n-1}$ and where (\ref{oscdensity}) would assume (\ref{traditional}) when we put
there the ``forbidden'' value $\alpha=2$ and $A=1$.
We emphasize that in a space of dimension $n$ the exponent due to the self-similar (non-local) Laplacian is always greater than the exponent due to the traditional (localized) Laplacian

\begin{equation}
 \label{selftradi}
\frac{2n}{\alpha}-1 > n-1 \,,\hspace{2cm} 0<\alpha=\delta-(n-1)<2
\end{equation}
It is quite remarkable that the self-similar Laplacian gives rise to unusual physical phenomena being qualitatively different from those found with a traditional Laplacian.

Whereas diffusion problems formulated with a traditional Laplacian describe traditional Gaussian statistics with finite variances,
diffusion problems formulated with Laplacian (\ref{laplaceddim}) describe random motions allowing long-range scale-free distributed jumps and yielding infinite variances referred to as L\'evi flights.
We devote the next section to this problem.

\section{Anomalous diffusion problem in $n$ dimensions - L\'evi flights}
\label{anodiff}

We consider an ensemble of particles of density $\rho({\bf r},t)$ where $\rho({\bf r},t){\rm d}^n{\bf r}$ denotes the number fraction of particles being located at a time $t$ in the volume element ${\rm d}^n{\bf r}$ which is attached to spacepoint ${\bf r}$. We refer to this picture as the ``diffusion picture''. The other picture of this model is a single randomly walking particle where $\rho({\bf r},t){\rm d}^n{\bf r}$ then means in this ``stochastic picture'' the probability to find the particle at time $t>0$ in the volume element ${\rm d}^n{\bf r}$ which is attached to spacepoint ${\bf r}$.
In what follows we describe a model of the space-time evaluation of the density $\rho$ where both pictures, the diffusion picture and the stochastic picture are possible physical interpretations. In the stochastic picture the present model describes a random walking particle which is walking continuously in time where the jump distance of the particle in any infinitesimal time interval $\delta t$ is
distributed according to a power law and isotropic in space. For such stochastic motions Mandelbrot coined the term {\it L\'evi flights}.
We will see that our above introduced self-similar Laplacian operator is the source of exactly this type of stochastic motion in the $n=1,2,3,..$-dimensional space.

We consider the problem for $t>0$.
We are especially interested in the characteristic asymptotic behavior for large times $t$.

The density to be analyzed is normalized for all times $t>0$ according to
\begin{equation}
 \label{densitynorm}
\int\rho({\bf r},t)\,{\rm d}^n{\bf r}=1
\end{equation}
We then define the diffusion problem by the following diffusion equation (in the stochastic picture Fokker Planck equation)

\begin{equation}
 \label{diffusioneq} \frac{\partial}{\partial t}\rho({\bf r},t)= -{\cal L}_{n,\delta}\rho({\bf r},t)
\end{equation}
where $-{\cal L}_{n,\delta}=\Delta_{n,\delta}$ denotes the Laplacian (\ref{laplaceddim}).
We further assume a prescribed initial distribution

\begin{equation}
\label{inidis}
\rho({\bf r},t=0)=\rho_0({\bf r})
\end{equation}
which is also normalized according to (\ref{densitynorm}).
We conceive the positive semi-definite operator ${\cal L}_{n,\delta}$ as the {\it diffusion generator}
where its eigenvalue spectrum is just the dispersion relation (\ref{disperela}) with the diffusional eigenmodes
$\phi_k({\bf r})=e^{i{\bf k}\cdot{\bf r}}$. The density $\rho({\bf r},t)$ writes in terms of eigenmodes
as Fourier transformation

\begin{equation}
 \label{eigendens}
\rho({\bf r,t})=\frac{1}{(2\pi)^n}\int {\hat \rho}({\bf k},t) e^{i{\bf k}\cdot{\bf r}}{\rm d}^n{\bf k}
\end{equation}
the Fourier amplitude ${\hat \rho}({\bf k},t)$ fulfills the evolution equation
\begin{equation}
 \label{amplieq}
\frac{\partial}{\partial t}{\hat \rho}({\bf k},t)=-\omega_{n,\delta}^2(k){\hat \rho}({\bf k},t)
\end{equation}
showing an exponential decay in time ($A=A_{n,\delta}>0$)

\begin{equation}
 \label{amlidudes}
{\hat \rho}({\bf k},t)= e^{-tA k^{\alpha}}{\hat \rho}_0({\bf k})
\end{equation}
where we have put $\alpha=\delta-(n-1)$ and ${\hat \rho}_0({\bf k})$ indicating the Fourier transform of the initial distribution $\rho_0({\bf r})$.
Unlike in the case of Gaussian diffusion, the non-locality of Laplacian (\ref{laplaceddim}) indicates that non-local particle jumps are admitted which are scale free distributed. We will come back to this important property more closely below.

We can also write the solution of (\ref{diffusioneq}) in the form

\begin{equation}
 \label{solform}
\rho({\bf r},t)= e^{-t{\cal L}_{n,\delta}}\rho_0({\bf r})
\end{equation}
which we will evaluate next. We observe that the normalization
of (\ref{solform}) is maintained at all times $t>0$

\begin{equation}
 \label{normalo}
\int \rho({\bf r},t){\rm d}^n{\bf r}=e^{-t{\cal L}_{n,\delta}}\int \rho_0({\bf r}) {\rm d}^n{\bf r}=1e^{-{\cal L}_{n,\delta}t} 1=1
\end{equation}
since $\rho_0({\bf r})$ is normalized with ${\cal L}_{n,\delta}1=0$ so that all powers higher than $m=0$ in the exponential
series of the time evolution operator $e^{-{\cal L}_{n,\delta}t}1$ applied on a constant yield vanishing contributions.
This indicates that the total particle number is a conserved quantity.
It is further illuminating to consider the diffusion processes more closely:
To this end we consider how the density $\rho({\bf r},t+\delta t)$ evolves from the density $\rho({\bf r},t)$
where $\delta t$ is an infinitesimal small time interval

\begin{equation}
 \label{generatoreq}
\rho({\bf r},t+\delta t)=e^{-\delta t{\cal L}_{n,\delta}}\rho({\bf r},t)
\end{equation}
where we can put $e^{-\delta t{\cal L}_{n,\delta}}\approx 1-\delta t{\cal L}_{n,\delta}$ and so

\begin{equation}
 \label{genrato}
\rho({\bf r},t+\delta t)=\rho({\bf r},t)-\delta t {\cal L}_{n,\delta}\rho({\bf r},t)
\end{equation}
The negative Laplacian ${\cal L}_{n,\delta}$ is the generator, generating the infinitesimal transformation
of the density from $t$ to $t+\delta t$.

Since $\rho({\bf r},t){\rm d}^n{\bf r}$ denotes the particle number fraction being at time $t$ in volume element
${\rm d}^n{\bf r}$ which is attached to the spacepoint ${\bf r}$, we can conceive the quantity

\begin{equation}
 \label{jumpbal}
\frac{\rho({\bf r},t+\delta t)-\rho({\bf r},t)}{\delta t}\sim \frac{\partial}{\partial t}\rho({\bf r},t)=
-{\cal L}_{n,\delta}\rho({\bf r},t)
\end{equation}
as the net balance of the particle number(fraction) departing and arriving in volume element ${\rm d}^n{\bf r}$ during the time interval
$\delta t$. In other words (\ref{jumpbal}) measures the number of particles jumping into the volume element minus the number of particles jumping out of the volume element.

The local net balance is due to the non-locality of the generating operator ${\cal L}_{n,\delta}\rho({\bf r},t)$ depending on all values of $\rho$ at time $t$ in the entire physical space $\R^n$ and
not (as in the case of Gaussian diffusion) only from the $\rho$-values in the local neighborhood.
We can express this local balance described by the diffusion equation (\ref{diffusioneq}) in terms of a continuity equation

\begin{equation}
 \label{continuity} \frac{\partial}{\partial t}\rho({\bf x},t)= -\nabla_{x}\cdot {\bf J}({\bf x},t)
\end{equation}
where we introduced the {\it particle
flux density ${\bf J}({\bf x})$} which we can write by using (\ref{vecfield}) in the form
\begin{equation}
\label{partflux}
{\bf J}({\bf x},t) = \frac{-1}{(\delta-(n-1))}\frac{h^{\alpha}}{2\zeta}\int \frac{{\bf r}}{r^{\delta+1}}\left\{\rho({\bf x}+{\bf r},t)-\rho({\bf x}-{\bf r},t)\right\}{\rm d}^n{\bf r} ,\,\hspace{1cm} 0 <\delta-(n-1) < 2
\end{equation}
which is determined up to an unimportant rotational gauge field (describing closed flux lines which do not change the local density (\ref{jumpbal})). We observe that equal distributions $\rho({ \bf x})=const$ would not
cause any particle flux. We can write the flux density also in the equivalent form
\begin{equation}
\label{partflux2}
{\bf J}({\bf x},t) = \frac{1}{(\delta-(n-1))}\frac{h^{\alpha}}{\zeta}\int \frac{{\bf r}}{r^{\delta+1}}\rho({\bf x}-{\bf r},t){\rm d}^n{\bf r} ,\,\hspace{1cm} 0 <\delta-(n-1) < 2
\end{equation}
where a further equivalent expression is obtained by exchanging ${\bf r}\rightarrow -{\bf r}$ in (\ref{partflux2}).
For $n=1$ these relations recover the expression found earlier \cite{michel2}.
We can conceive (\ref{partflux}) or equivalently (\ref{partflux2}) as the (non-local) constitutive law connecting particle flux and density replacing the (local) Fick's law of Gaussian diffusion.

Let us consider the particle jump rate into a small volume element $\delta V$ around ${\bf r}=0$ due to a
localized distribution $Q({\bf x},t=0)= \delta^n({\bf x})$. This localized particle distribution induces at
spacepoint of distance $x=|{\bf x}|$ instantaneously the flux (due to non-local particle jumps of distance $x$)

\begin{equation}
\label{fluxflux}
{\bf J}({\bf x},t=0)=\frac{h^\delta}{(\delta-(n-1))\zeta}\frac{\bf x}{x^{\delta +1}}
\end{equation}
These non-local particle jumps due to the flux (\ref{fluxflux}) must cause at spacepoint ${\bf x}$ a particle
balance which must be positive due to counting particle jumps from ${\bf r}=0$ to ${\bf x}$, i.e. jumps over a distance $x$. We obtain with (\ref{fluxflux})

\begin{equation}
 \label{partbalan}
\frac{\partial}{\partial t}Q({\bf x},t=0)=-\nabla_{x}\cdot{\bf J}({\bf x},t=0)=\frac{h^{\alpha}}{\zeta}\, x^{-\delta-1} >0\,,\hspace{1cm} \forall {\bf x} \neq 0
\end{equation}
which is positive and scaling as $\sim x^{-\delta-1}$ and decays always with distance $x$ whatever the dimension $n$ since $n<\delta+1<n+2$ and is nonzero in the entire space whatever the jumping distance $x$ of the particles. The relation (\ref{partbalan}) holds everywhere {\it except} in the origin ${\bf x}=0$. Due to the stochastic spatial isotropy of the particle jumps, integration
of (\ref{partbalan}) over the unit sphere gives the jump rate of all jumps with distance $R$ at $t=0+$. This rate is given by

\begin{equation}
 \label{ratex}
\ds -\int_{|{\bf n}=1|}{\rm J}\cdot {\bf n}R^{n-1}{\rm d}\Omega({\bf n})= \frac{2\pi^{\frac{n}{2}}}{\Gamma(\frac{n}{2})} R^{n-1}\frac{h^{\alpha}}{\zeta}\, R^{-\delta-1} \sim R^{-\alpha-1} \,,\hspace{0.5cm} 0<\alpha=\delta-(n-1)<2 , \hspace{1cm} R\neq 0
\end{equation}
where $\frac{2\pi^{\frac{n}{2}}}{\Gamma(\frac{n}{2})}R^{n-1}$ is the surface of the sphere of radius $R$.
We can interpret (\ref{ratex}) as follows: The probability that a particle which is at $t=0$ located in the origin ${\bf r}=0$ undertakes a jump of distance $R$ within the infinitesimal time interval $\delta t$ scales as $R^{-\alpha-1}$.
It follows that the jump rate of all jumps of distance $R$ at $t=0+$ decays spatially scaling as $\sim R^{-\alpha-1}$ with the jump distance $R$ and indeed is L\'evi distributed
where $0<\alpha<2$ is the band of admissible L\'evi-parameter $\alpha$ {\it whatever the dimension $n$ of the physical space}.

A crucial rule plays the space-time representation of the time evolution operator from which we considered the small time regime in (\ref{partbalan}). This propagator is generally defined by

\begin{equation}
 \label{evalprop}
Q({\bf r},t)=e^{-t{\cal L}_{n,\delta}}\delta^n({\bf r})
\end{equation}
where $\delta^n({\bf r})$ denotes the $n$-dimensional Dirac's $\delta$-function.
In the stochastic picture the interpretation is as follows: The kernel $Q({\bf r},t)$ describes then the {\it conditional probability density} and $Q({\bf r},t){\rm d}^n{\bf r}$ denotes the probability to find the particle which was located at $t=0$ in the origin ${\bf r}=0$ at time $t$ in the volume element ${\rm d}^n{\bf r}$ attached to
the spacepoint ${\bf r}$.
Correspondingly $Q({\bf r},t)$ fulfills then the initial condition
\begin{equation}
 \label{inicond}
Q({\bf r},t=0)=\delta^n({\bf r})
\end{equation}
and is hence itself a normalized probability distribution solving (\ref{diffusioneq}). We deduce some integral relations thereof in the appendix \ref{appendix1}.
The density (\ref{solform}) can then be represented by the convolution

\begin{equation}
 \label{propa}
\rho({\bf r},t)=\int Q({\bf r}-{\bf r}',t)\rho_0({\bf r}'){\rm d}^n{\bf r}'
\end{equation}
Taking into account that

\begin{equation}
 \label{deltafun}
\delta^n({\bf r})=\frac{1}{(2\pi)^n}\int e^{i{\bf k}\cdot{\bf r}}{\rm d}^n{\bf k}
\end{equation}
together with the property

\begin{equation}
 \label{eigenpropn}
{\cal L}_{n,\delta}^m e^{i{\bf k}\cdot{\bf r}}= \omega_{n,\delta}^{2m}(k) e^{i{\bf k}\cdot{\bf r}},\,\hspace{1cm} m=0,1,2,..\in \N
\end{equation}
where $\omega_{n,\delta}^{2}(k)$ denotes the dispersion relation (\ref{disperela}), we can write
for any sufficiently smooth function $f(\xi)$ the relation
$f(t{\cal L}_{n,\delta})e^{i{\bf k}\cdot{\bf r}}= f(t\omega_{n,\delta}^{2}(k)) e^{i{\bf k}\cdot{\bf r}}$.
By using this property for the exponential operator $e^{-{\cal L}_{n,\delta}t}$
we obtain $Q({\bf r},t)$ in the form

\begin{equation}
 \label{exprel}
Q(r,t)=\frac{1}{(2\pi)^n}\int e^{-A_{n,\delta}k^{\alpha} t} e^{i{\bf k}\cdot{\bf r}}{\rm d}^n{\bf k}
\end{equation}
where $0<\alpha=\delta-(n-1)<2$. The kernel is spatially isotropic and depends only on $r=|{\bf r}|$ due to the isotropic
symmetry of the $\delta$-function. We note that the linear order in $t$ of this Fourier integral coincides with
(\ref{partbalan}) constituting the regime of small times $t$ (appendix \ref{appendix1}).

Distributions of the form (\ref{exprel}) are referred to as {\it L\'evi distributions}
\cite{levi,mandel2}. In contrast to Gaussian distributions, L\'evi-distributions exhibit diverging mean fluctuations (all even moments are diverging). This can be directly verified from

\begin{equation}
\label{evenmo}
<r^2>=\left\{\int Q(r,t)r^2e^{-i{\bf k}\cdot{\bf r}}{\rm d}^n{\bf r}\right\}_{k=0}= -\nabla_{k}\cdot\nabla_{k}{\tilde Q}(k,t)|_{k=0} \rightarrow \infty
\end{equation}
which is fulfilled by ${\tilde Q}(k,t)=e^{-A_{n,\delta}k^{\alpha} t}$ for positive $\alpha$ in the interval $0<\alpha<2$. It is interesting to see that the condition of existence of the Laplacian (\ref{laplaceddim}) leads to the same admissible $\alpha$-band $0<\alpha<2$ as the condition of divergence of the variance (\ref{evenmo}), equivalent with the condition of non-differentiability of ${\tilde Q}(k,t)$ at $k=0$.

All odd moments are vanishing due to the isotropic symmetry of the distribution $Q(r,t)$.
For the further evaluation it is convenient to introduce the function defined by the surface integral over the unit sphere

\begin{equation}
\label{spheric}
G_n(\tau)=\frac{1}{(2\pi)^n}\int_{|{\hat k}|=1}{\rm d}\Omega({\hat k})e^{\tau {\hat k}_1} =\frac{1}{(2\pi)^n}\int \delta(k-1)e^{\tau k_1}{\rm d}^n{\bf k}
\end{equation}
where $k_1$ denotes any Cartesian component of the unit vector ${\hat k}$.
We observe due to the spherical symmetry that
$G_n(\tau)$ contains only even powers in $\tau$, i.e. only the even part $\cosh{\tau{\hat k}_1}$ of the integrand contributes to (\ref{spheric}). The surface integral (\ref{spheric}) is evaluated explicitly in appendix
\ref{appendix1} (eq. (\ref{Gnclosed})).
The kernel (\ref{exprel}) can be further written as

\begin{equation}
 \label{exprel2}
Q(r,t)=r^{-n}P(\frac{A_{\delta,n}t}{r^{\alpha}})
\end{equation}
where $P$ is a function of the ``scaled time'' $\lambda^{\alpha}=\frac{A_{\delta,n}t}{r^{\alpha}}$ only.
In the following
we keep in mind that the constant $A_{\delta,n}$ depends on $n$ and $\delta$ and skip these subscripts.
The function $P$ takes then the form
\begin{equation}
\label{Pkernel}
P(\lambda)=\int_0^{\infty} e^{-\lambda^{\alpha}\xi^{\alpha}} G_n(i\xi)\,\xi^{n-1}{\rm d}\xi
\end{equation}
To consider large $\lambda^{\alpha}=\frac{A t}{r^{\alpha}}>>1$ we can write by introducing the
``fast'' variable $u=\lambda\xi$, and we emphasize that $0<\alpha<2$ and hence
$\frac{1}{\alpha}$ is a positive exponent, so we can rewrite

\begin{equation}
\label{Pass}
P(\lambda) = \lambda^{-n}\int_0^{\infty}e^{-u^{\alpha}}u^{n-1}G_n(i\lambda^{-1}u){\rm d}u
\end{equation}
and hence the kernel $Q=r^{-n}P$ has the form

\begin{equation}
 \label{Qrt}
Q(r,t)=\frac{1}{(At)^{\frac{n}{\alpha}}}W\left(\frac{r^{\alpha}}{(At)}\right) \,,\hspace{2cm} 0<\alpha=\delta-(n-1)<2
\end{equation}

\begin{equation}
 \label{Wint}
W(\lambda^{-\alpha})=\int_0^{\infty}e^{-u^{\alpha}}u^{n-1}
G_n(i\lambda^{-1}u){\rm d}u \,,\hspace{1cm} \lambda=\frac{(At)^{\frac{1}{\alpha}}}{r}
\end{equation}
and the function $G_n$ depends only on the dimension $n$ of physical space and is defined by (\ref{spheric}). We note that equations (\ref{Qrt}) with (\ref{Wint}) are exact relations.
Generally, except in certain cases to be considered, the integral (\ref{Wint}) cannot be obtained in closed form.
Despite $\alpha=2$ is a forbidden case in our model, we can formally consider this case for which (\ref{Qrt}) with (\ref{Wint}) can be evaluated in closed form and lead to a Gaussian distribution

\begin{equation}
\label{gaussian}
Q_{g}(r,t)= \frac{1}{(4\pi At)^{\frac{n}{2}}}e^{\ds -\frac{r^2}{4At}}
\end{equation}
where here $A=A_2$ which is not defined by our model as dispersion relation (\ref{dispo}) is diverging for $\alpha=2$.
Representation (\ref{Qrt}) is in a sense analogue to the
Gaussian distribution (\ref{gaussian}) where the time dependence of the normalization
factor of (\ref{Qrt}) is given by $(At)^{-\frac{n}{\alpha}}$ and exhibits in the Gaussian case
($\alpha=2$) $(At)^{-\frac{n}{2}}$ and leading to (\ref{gaussian}).
\newline\newline
\noindent {\bf Asymptotic regime $\lambda^{\alpha}=\frac{(At)}{r^{\alpha}} >>1$} where $\lambda^{\alpha}$ denotes a scaled time: In this regime, relation (\ref{Wint}) assumes asymptotically the form

\begin{equation}
 \label{assrel}
\begin{array}{l}
\ds Q(r,At>>r^{\alpha}) \sim \frac{G_n(0)}{(At)^{\frac{n}{\alpha}}}\left(I(n,\alpha)-\frac{r^2}{(At)^{\frac{2}{\alpha}}}I(n+2,\alpha)+..O(\frac{r}{(At)^{\frac{1}{\alpha}}})^4\right) \nonumber \\ \ds \nonumber \\
\ds \hspace{1cm} \sim Q(t)=\frac{G_n(0)(I(n,\alpha)}{(At)^{\frac{n}{\alpha}}} \rightarrow 0
\end{array}
\end{equation}
which decays in the dominant term in time as $t^{-\frac{n}{\alpha}}$ independent on $r$ and where
\begin{equation}
\label{In}
I(n,\alpha)=\int_0^{\infty}e^{-u^{\alpha}}u^{n-1}{\rm d}u
\end{equation}
and
\begin{equation}
 \label{Gn0}
G_n(0)= \frac{2}{(4\pi)^{\frac{n}{2}}\Gamma(\frac{n}{2})}
\end{equation}

The asymptotic relation (\ref{assrel}) describes the manner how the distribution approaches the
thermodynamic Boltzmannian equal distribution. Since (\ref{assrel}) does not contain any information about the initial positions of the particles we can conclude that the leading term $Q(t)$ in (\ref{assrel}) is universal and holds for any initial distribution
$\rho_0({\bf r})$. This follows also from the asymptotic relation where we assume that $At/R^{\alpha}>>1$ and
$R>>1$: The leading contribution to the density in this regime then is (with $Q({\bf r},t)\sim Q(t)$)

\begin{equation}
\label{assrelaforall}
\rho({\bf r},t)=\int Q({\bf r}-{\bf r}',t)\rho_0({\bf r'}){\rm d}^n{\bf r'} \sim Q(t)\int \rho_0({\bf r'}){\rm d}^n{\bf r'}=Q(t)
\end{equation}
which leads indeed for {\it any initial distribution} $\rho_0({\bf r})$ to the {\it identical spatially uniform asymptotics} $Q(t) \sim t^{-\frac{n}{\alpha}}\rightarrow 0$ where the exponent depends only on the ratio $\frac{n}{\alpha}$ of the physical dimension $n$ and the L\'evi parameter $\alpha$ and with the restriction $\frac{n}{2}<\frac{n}{\alpha}<\infty$.
If $t$ exceeds all limits the system of diffusing particles ``forgets'' its past and approaching all the same thermodynamic attractor $Q(t)\rightarrow 0$. However, since the scaled time $At/r^{\alpha}$ is for finite $t$ never
a large quantity in the entire space as there is always a region where $At/r^{\alpha} << 1$ is still small and hence (\ref{assrel}) is not (yet) valid there. In other words: The spherical region for which
the spatially quasi-equal distribution $Q(t)$ of (\ref{assrel}) is valid and hence which is close to thermodynamic equilibrium is expanding in time slower than $r(t)\sim (At)^{\frac{1}{\alpha}}$. The volume of this expanding region scales as $V(t)\sim r^n(t)\sim (At)^{\frac{n}{\alpha}}$. The asymptotic uniform distribution $Q(t)\sim 1/V(t)\sim (At)^{-\frac{n}{\alpha}}$ scales as the inverse of the volume $V(t)$ of thermodynamic quasi-equilibrium where $Q(t)$
is the {\it quasi-equal distribution} within this region of ``quasi-equilibrium''. The universal scaling behavior of the type (\ref{assrel}) in fractional models of anomalous diffusion was already noted earlier \cite{metzler}.

For all dimensions $n$ of the physical space holds: The algebraic decay of the kernel $Q(r,t) \sim Q(t)$ being independent from $r$ for large scaled times $At/r^{\alpha}>>1$ is a necessary consequence of Boltzmann's postulate saying that the location of a diffusing particle becomes completely undetermined without any preferred location in the thermodynamic equilibrium indicating the complete ``loss of information'' about the location of the particle. This corresponds to the thermodynamic necessary condition of approaching maximum (infinite) entropy.
We will demonstrate this by the following brief consideration of the entropy which we define here as negative H-function \cite{lipshitz}

\begin{equation}
 \label{entropy}
S(t) \sim -\int Q(r,t)\log{Q(r,t)}\,{\rm d}^n{\bf r}
\end{equation}
We are especially interested in this quantity in the large times limit $t\rightarrow \infty$.
In this regime we can put

\begin{equation}
 \label{entropyassymp}
S(t) \sim -\log\{(At)^{-\frac{n}{\alpha}}\}\int Q(r,t)\,{\rm d}^n{\bf r} \sim \frac{n}{\alpha}\log(t) \rightarrow \infty
\end{equation}
which diverges logarithmically and spatially homogenously in time indicating in the stochastic picture that the state of complete uncertainty about the
location of the particle is approached as the global time $t$ tends to infinity. It follows that (\ref{entropyassymp}) is universal and independent from the initial distribution $\rho_0({\bf r})$.
In other words relation (\ref{entropyassymp}) expresses
the validity of the H-theorem due to Boltzmann which is equivalent to the fact that distributions are broadening in time approaching equal distribution as time tends towards infinity.
The divergence of the maximum entropy in (\ref{entropyassymp}) is reflecting the fact that the available volume for any particle becomes infinite as $t\rightarrow\infty$ whereas the particle number remains a finite constant. The entropy would approach a finite equilibrium value if for $t\rightarrow\infty$ the volume accessible for the particles would be finite. A finite volume, however would be in contradiction to the self-similarity and as a consequence non-locality of the Laplacian.
For a ``real'' physical system in nature therefore, self-similarity can only be approximatively fulfilled.

Let us briefly consider the one-dimensional case $n=1$: There we have $\alpha=\delta-(n-1)=\delta$ with

\begin{equation}
 \label{asn1}
Q_1(r,At>>r^{\delta}) \sim \frac{I(1,\delta)}{\pi}(A t)^{-\frac{1}{\delta}}\,,\hspace{2cm} 0<\delta<2
\end{equation}
and

\begin{equation}
 \label{I1delta} I(1,\delta)= \int_0^{\infty}e^{-u^{\delta}}{\rm d}u
\end{equation}

Expressions (\ref{asn1}) with (\ref{I1delta}) is in accordance with the asymptotic relation obtained in \cite{michel2} for the one-dimensional case.
There is one single case remaining, where $Q(r,t)$ can be obtained in closed form: From (\ref{Wint}) that
this is the case for $\alpha=n$ which can be only fulfilled for $n < 2$. Since $n\in \N$ the only case is $n=\delta=1$. We obtain then

\begin{equation}
\label{G1}
 G_1(i\xi)=\frac{1}{\pi}\cos{\xi}
\end{equation}
and (\ref{In}) yields
\begin{equation}
 \label{Wd1}
W(\lambda^{-1})=\frac{1}{\pi}\int e^{-u}\cos{(\lambda^{-1}u)}\,{\rm d}u=\frac{1}{\pi}\frac{(At)^2}{(At)^2+r^2}
\end{equation}
The distribution (\ref{Qrt}) is then obtained as ($r=|x|$)
\begin{equation}
\label{cauchy}
Q(r,t)=\frac{1}{\pi}\frac{At}{(At)^2+x^2}
\end{equation}
which is known as Cauchy distribution \cite{mandel2} which has also the property of diverging even moments.
The Cauchy distribution hence is the result of a L\'evi flight for a special L\'evi parameter $\alpha=\delta=1$. Sometimes this motion is referred to as Cauchy flight. The Cauchy case stands out in the present model and appears uniquely for $n=\delta=1$.
The Cauchy distribution behaves for large $At >> |x|$ as $Q(t) \sim \frac{1}{\pi}(At)^{-1}$ in accordance with
(\ref{assrel}) for $\alpha=n=1$ (with $I(1,1)=1$ and $G_1(0)=\frac{1}{\pi}$).

\section{Conclusions}

We have deduced a Laplacian operator for a $n$-dimensional infinite space
with self-similar symmetry. The Laplacian introduced in this paper fulfils all criteria of ellipticity: linearity, self-adjointness, isotropic symmetry, negative semi-definiteness, and translational symmetry. The self-similar symmetry makes this Laplacian operator non-local. We proved that this self-similar Laplacian coincides (up to a strictly positive prefactor) to the {\it fractional Laplacian}. To have a physical picture, the self-similarity can be conceived as an elastic medium with self-similar scaling invariant harmonic interparticle interactions.
Employing this picture we have deduced the dispersion relation with the density of normal oscillator modes.
The density of normal modes in the frequency space fulfills a scaling law $\sim \omega^{\frac{2n}{\alpha}-1}$ with a characteristic strictly positive exponent being always greater than $n-1$ which would be the exponent due to a traditional Laplacian of the $n$-dimensional space.

We analyzed a diffusion problem defined by a Fokker-Planck equation by
employing the self-similar Laplacian. The model describes anomalous diffusion
allowing non-local particle jumps where the jump distances are L\'evi-distributed (L\'evi flights). The solutions of the Fokker Planck equation are L\'evi-stable distributions with characteristic algebraic decay $\sim t^{-\frac{n}{\alpha}}$ in the regime of large scaled times with the L\'evi parameter $\alpha$ being in the interval $0<\alpha<2$ whatever the dimension $n$. The present model of anomalous diffusion could have some applications on physical phenomena dominated by non-Brownian erratic motions such as for instance in turbulence. We hope the present approach inspires further work in such directions.

Moreover, the present approach can be used as a starting point to describe continuous field problems governed by self-similar constitutive laws in statics and dynamics. The physical nature of the fields can be of any kind, such as of mechanical or electromagnetical nature.
A corresponding self-similar electromagnetic field theory will be presented in a sequel.

\section{Acknowledgement}
This paper is dedicated to the memory of our dear friend Prof. Arne Wunderlin (Stuttgart, Germany).

\section{Appendix}
\label{appendixA}

\subsection{Derivation of the vector field (\ref{vecfield})}
\label{gausstheo}

Here we deduce the vector field (``dielectric displacement'') ${\bf D}$ of relation (\ref{vecfield}). To this end
we start with the Laplacian (\ref{laplaceddim})

\begin{equation}
 \label{laplaceddim2}
\ds \Delta_{(n,\delta)}u({\bf x})= \frac{h^{\alpha}}{2\zeta}\int \frac{\left(u({\bf x}+{\bf r})+u({\bf x}-{\bf r}) -2u({\bf x})\right)} {r^{\delta+1}}{\rm d}^n{\bf r} ,\,\hspace{1cm} 0 <\delta-(n-1) < 2
\end{equation}
and use the identity

\begin{equation}
\label{iden}
r^{-\delta-1}=\frac{-1}{\delta-(n-1)}\nabla_{\bf r}\cdot(r^{-\delta-1}{\bf r})
\end{equation}
with $\nabla \cdot ({\bf a}b)=b\nabla\cdot{\bf a}+{\bf a}\cdot\nabla b$. We note that $0<\delta-(n-1)<2$ is always non-zero.
By applying the Gauss theorem for the boundary integral term we get

\begin{equation}
 \label{iden2}
\begin{array}{l}
\ds \int r^{-\delta-1}\left(u({\bf x}+{\bf r})+u({\bf x}-{\bf r}) -2u({\bf x})\right){\rm d}^n{\bf r}= \nonumber \\ \nonumber \\
\ds \lim_{R\rightarrow\infty}\int_{\partial V(R)}r^{n-1}{\rm d}\Omega({\bf n})\cdot\left\{\frac{-1}{\delta-(n-1)}\cdot(r^{-\delta})\left(u({\bf x}+{\bf r})+u({\bf x}-{\bf r}) -2u({\bf x})\right)\right\}+ \nonumber \\ \nonumber \\ \ds +\frac{1}{\delta-(n-1)}\int(r^{-\delta-1}{\bf r})\cdot\nabla_{r}\left(u({\bf x}+{\bf r})+u({\bf x}-{\bf r}) -2u({\bf x})\right){\rm d}^n{\bf r}
\end{array}
\end{equation}
The boundary integral over $\partial_{V(R)}$ scales as $R^{-(\delta-(n-1))}\rightarrow 0$ ($0<\delta-(n-1)<2$) and is hence vanishing as $R\rightarrow\infty$ ($\delta-(n-1)>0$). When we further use $\nabla_{r}\left(u({\bf x}+{\bf r})+u({\bf x}-{\bf r}) -2u({\bf x})\right)=\nabla_{\bf x}\left(u({\bf x}+{\bf r})-u({\bf x}-{\bf r})\right)$
we can write (\ref{laplaceddim2}) as a divergence

\begin{equation}
 \label{nalbarep}
\ds \Delta_{(n,\delta)}u({\bf x})=\nabla_{\bf x}\cdot{\bf D}
\end{equation}
with the vector field ${\bf D}$ being determined (up to a unimportant rotational field)

\begin{equation}
 \label{vecS}
{\bf D}({\bf x})=\frac{1}{\delta-(n-1)}\frac{h^{\alpha}}{2\zeta}\int r^{-\delta-1}{\bf r}\left\{u({\bf x}+{\bf r})-u({\bf x}-{\bf r})\right\}{\rm d}^n{\bf r} \,,\hspace{1cm} 0<\delta-(n-1)<2
\end{equation}
which is relation (\ref{vecfield}). We observe that the integrand is an even function of ${\bf r}$. Taking into account that the volume integral over any odd function of ${\bf r}$ is vanishing
we can also write

\begin{equation}
 \label{alsoS}
{\bf D}({\bf x})=\frac{1}{\delta-(n-1)}\frac{h^{\alpha}}{\zeta}\int \frac{{\bf r}} {r^{\delta+1}}
u({\bf x}+{\bf r})
{\rm d}^n{\bf r} \,,\hspace{1cm} 0<\delta-(n-1)<2
\end{equation}
or equivalently (by replacing ${\bf x}+{\bf r} \rightarrow {\bf r}$)

\begin{equation}
 \label{alsoSb}
{\bf D}({\bf x})=-\frac{1}{\delta-(n-1)}\frac{h^{\alpha}}{\zeta}\int
\frac{{\bf x}-{\bf r}}{|{\bf x}-{\bf r}|^{\delta+1}}
u({\bf r})
{\rm d}^n{\bf r} \,,\hspace{1cm} 0<\delta-(n-1)<2
\end{equation}
where only the even part of the integrand namely (\ref{vecS})) contributes. All representations
(\ref{vecS}) and (\ref{alsoS}) or (\ref{alsoSb}) exist in the band $0<\delta-(n-1)<2$ just as the Laplacian (\ref{laplaceddim2}).
(\ref{alsoS}) can be further written in the form of a gradient of a scalar potential

\begin{equation}
 \label{pot}
{\bf D}({\bf x})=\nabla_{\bf x}\Phi({\bf x})
\end{equation}
where the scalar potential $\Phi({\bf x})$ can be written as

\begin{equation}
 \label{poPhi}
\Phi({\bf x})= \frac{h^{\alpha}}{(\delta-(n-1))(\delta-1)\zeta}\int |{\bf x}-{\bf r}|^{1-\delta} u({\bf r}){\rm d}^n{\bf r} \,\hspace{1cm} \delta \neq 1 \hspace{0.5cm} 0<\delta-(n-1)<2
\end{equation}

For $n=1$ this expression recovers those obtained in our previous paper (\cite{michel}).
We note that the scalar $\Phi({\bf x})$ is a convolution of the scalar field $u$ with the convolution kernel being the
power function (``Riesz potential'') $|{\bf r}-{\bf x}|^{1-\delta}$ where $n-1<\delta<n+1$. The special case of $\delta=1$ is appears only for $n=1$. For $n\geq 2$ we have $1\leq n-1 <\delta<n+1$ and hence $\delta >1$ for dimensions $n=2,3,..$. In these cases $R^{1-\delta}$ is vanishing at infinity and singular at $R=0$. Only the case of one dimension $n=1$ with $0<\delta<2$ has for $0<\delta<1$ the ``anomaly'' that the Riesz potential is diverging at infinity and vanishing in the origin.
It follows that

\begin{equation}
 \label{tradlaplaceeq}
\Delta_{n,\delta}u({\bf x})=\nabla_{\bf x}\cdot \nabla_{\bf x}\Phi({\bf x})
\end{equation}
where $\nabla_{\bf x}\cdot \nabla_{\bf x}$ denotes the traditional Laplacian of the $n$-dimensional space.
Relation (\ref{tradlaplaceeq}) recovers for $n=1$ the expression we obtained earlier \cite{michel}.

The case $\delta=1$ which can only occur for $n=1$ furthermore again stands out among all the others. This case e.g. lead us to the Cauchy-distribution in section \ref{anodiff}.
Evaluation of (\ref{poPhi}) for $n=\delta=1$ yields

\begin{equation}
 \label{Phi11}
\Phi_{\delta=1,n=1}({\bf x})=-\frac{h}{\zeta}\int_{-\infty}^{\infty}\ln(|x-\tau|)u(\tau){\rm d}\tau
\end{equation}
which has been already obtained in \cite{michel}.

\subsection{Determination of $A_{n,\delta}$}
\label{appAdeltan}

In this appendix we determine the constant $A_{n,\delta}=\omega_{n,\delta}^2(k=1)$ occurring in the dispersion relation (\ref{disperela}) $\omega_{n,\delta}^2(k)=A_{n,\delta}k^{\alpha}$ ($\alpha=\delta-(n-1)$).
We introduce ${\bf r}\cdot{\bf k}=krn_1$, (${\bf n}\cdot{\bf k}=kn_1$).
Then we have to evaluate (\ref{dispo})
\begin{equation}
 \label{Aconstant}
\omega^2(k=1)_{n,\delta}=A_{n,\delta}= \frac{1}{2}J_{\alpha}A_{1,\alpha}
\end{equation}
with
\begin{equation}
 \label{eval1}
A_{1,\alpha}=\omega_{n=1,\delta=\alpha}^2(k=1)=\frac{2h^{\alpha}}{\zeta}\int_0^{\infty}\frac{(1-\cos(\tau))}{\tau^{\alpha+1}}{\rm d}\tau \,\hspace{2cm} 0<\alpha=\delta-(n-1)<2
\end{equation}

In view of (\ref{Aconstant}) we can conclude that the dispersion relation of the $n$-dimensional space corresponds
to that one in one dimension when we replace $\delta\rightarrow \alpha$ up to a prefactor $\frac{J_{\alpha}}{2}$ which depends only on the exponent $\alpha$ and dimension $n$ of the physical space.
The constant $A_{1,\alpha}$ of (\ref{eval1}) is nothing but the {\it dispersion relation for the one-dimensional case} $\omega_{n=1,\delta}^2(k=1)$ which we already deduced in the one-dimensional model \cite{michel2}. For the evaluation we refer to that paper.
We obtained there

\begin{equation}
 \label{A1delta}
A_{1,\alpha}=\frac{h^{\alpha}\pi}{\zeta \alpha ! \sin{\frac{\pi\alpha}{2}}} >0 \,,\hspace{2cm} 0<\alpha=\delta-(n-1) <2
\end{equation}
where this constant is strictly positive and finite in the admitted $\alpha$-range.
In (\ref{Aconstant}) occurs a surface integral on the unit-sphere
\begin{equation}
\label{Jint}
\ds J_{\alpha}=\int_{|{\bf n}|=1}{\rm d}\Omega({\bf n})|n_1|^{\alpha} = \frac{2\pi^{\frac{n-1}{2}}\Gamma(\frac{\alpha+1}{2})}{\Gamma(\frac{\alpha+n}{2})},\hspace{2cm} 0<\alpha=\delta-(n-1)<2
\end{equation}

This surface integral is straight-forwardly evaluated from the integral 
$I_{\alpha}^{(n)}=\int e^{-r^2}|x|^{\alpha}{\rm d}^n{\bf r}$ where $x=rn_1$ and by seperating the surface integration and the
radial integration and by decomposing it into a product of $n$ integrals over the Cartesian coordinates. After some cumbersome
algebra by using the doublication formula together with Euper reflection formulas for $\Gamma$-functions (see Abramovitz \& Stegun \cite{abramowitz}, page 256, formulas 6.1.18 where $2z=\alpha+1$ and 6.1.17 with $z=\frac{\alpha}{2}$), one finally arrives for the constant (\ref{Aconstant}) at the compact expression

\begin{equation}
 \label{final}
A_{n,\delta}=A_{n,\alpha} = \frac{h^{\alpha}}{\zeta} \frac{\pi^{\frac{n}{2}}\Gamma(1-\frac{\alpha}{2})}{2^{\alpha-1}\alpha\Gamma(\frac{\alpha+n}{2})} ,\hspace{1cm}  0<\alpha<2
\end{equation}
which is in accordance with the normalization constant known from literature, e.g. in \cite{vazquez} (and see the references therein) which occurs by defining the (negative) {\it fractional Laplacian}
$(-\Delta)^{\frac{\alpha}{2}}$ being defined as the operator having the Fourier transform $k^{\alpha}$ with $0<\alpha<2$. It follows from the scaling behavior of the dispersion relation $\omega^2(k)_{n,\delta}=A_{n,\delta}k^{\alpha}$, that the fractional Laplacian $-(-\Delta)^{\frac{\alpha}{2}}$ and self-similar Laplacian defined by (\ref{laplaceddim}), (\ref{laplaceddimb}) are linked by the relation

\begin{equation}
 \label{fractional}
\Delta_{(n,\delta)}=-A_{n,\alpha}(-\Delta)^{\frac{\alpha}{2}}, \hspace{1cm}  0<\alpha<2
\end{equation}
with the strictly postive prefactor $A_{n,\alpha}$ which is determined by (\ref{final}) for any dimension $n=1,2,3,..$ of the physical space. We emphasize that the strict positiveness of $A_{n,\alpha}$ within $0<\alpha<2$ indeed guarantees the ellipticity of $\Delta_{(n,\delta)}$ represented by (\ref{laplaceddim}), (\ref{laplaceddimb}).
\\ \\
Let us consider briefly $n=1$: Then $\alpha=\delta$ and $J_{\delta}=2$ and so (\ref{Aconstant}) yields then
our previously deduced expression of \cite{michel2}
\begin{equation}
\label{prexp}
A_{1,\delta}=\frac{h^{\delta}\pi}{\zeta \delta ! \sin{\frac{\pi\delta}{2}}} >0 \,,\hspace{2cm} 0<\delta<2
\end{equation}

Let us also evaluate $n=3$: $\alpha=\delta-2$ and the integral (\ref{Jint}) yields then

\begin{equation}
 \label{alphdelm2}
J_{\alpha=\delta-2}= \frac{4\pi}{\delta-1}
\end{equation}
and hence
\begin{equation}
 \label{Aalphn3}
A_{3,\delta}=\frac{2h^{\delta-2}\pi^2}{\zeta(\delta-1)!\sin{\frac{\pi(\delta-2)}{2}}} >0\,,\hspace{2cm} 2<\delta<4
\end{equation}

In all cases the necessary condition $\omega^2(1) >0$ which follows already from the positiveness of the integrand of (\ref{eval1}).

\section{Some useful integrals}
\label{appendix1}

In this appendix we deduce a rather remarkable relation when we take into account the following identities, namely those of equations (\ref{partbalan}) and (\ref{evalprop}) being the linear order in $t$ of relation (\ref{exprel}). First we have

\begin{equation}
 \label{Qt}
\frac{\partial Q}{\partial t}(r,t=0)=\Delta_{n,\delta}\delta^n({\bf r})=
-\frac{A_{n,\alpha}}{(2\pi)^n}\int e^{i{\bf k}\cdot{\bf r}}k^{\alpha}{\rm d}^n{\bf k} =\frac{h^{\alpha}}{\zeta}\int \frac{\delta^n({\bf r}-{\bf r}')}{{r'}^{\alpha+n}}{\rm d}^n{\bf r}'=\frac{h^{\alpha}}{\zeta}r^{-\alpha-n} 
\end{equation}
where the latter integral is obtained by application of the Laplacian on the $\delta$-function for ${\bf r}\neq 0$ and yields (\ref{partbalan}).
The Fourier integral can be rewritten for $r\neq 0$

\begin{equation}
 \label{QTb}
\frac{\partial Q}{\partial t}(r,t=0)=-\frac{A_{n,\alpha}}{r^{\alpha+n}}\int_0^{\infty}G_n(i\tau)\tau^{\delta}{\rm d}\tau
\end{equation}
where $G_n$ is the surface integral defined in (\ref{spheric}) which is evaluated in closed form below.
Comparision with the explicit expression (\ref{partbalan}) yields the integral relation

\begin{equation}
 \label{integrel}
\int_0^{\infty} G_n(i\tau)t^{\delta}{\rm d}\tau= -\frac{h^{\alpha}}{\zeta A_{n,\alpha}} < 0
\end{equation}
which can be explicitly confirmed by evaluating the left hand side for $n=1$ and $n=3$.\newline\newline
\noindent {\bf Surface integrals}

We choose a Cartesian coordinate system with and ${\bf n}=(n_1,..,n_n)^{tr}$ with ${\bf n}\cdot{\bf n}=1$ being a parameterization of the $n$-dimensional unit-sphere. Let us then evaluate the integral of an integer power of any Cartesian coordinate

\begin{equation}
 \label{integm}
{\cal A}^{(n)}_m=\int_{|{\bf n}|=1}{\rm d}\Omega({\bf n})\,n_1^{m}\,,\hspace{1cm} m=0,1,2,..\in \N_0
\end{equation}
where we integrate over the surface of the $n$-dimensional unit-sphere. The superscript $(..)^{(n)}$ indicates the dimension ($n=1,2,3,.\in \N$) of the space. We observe that
all integrals of odd powers (odd functions) ${\cal A}_{2m+1}^{(n)}=0$ are vanishing.
Only integrals over even powers
are non-vanishing and by applying the Gauss-theorem we find the recursion

\begin{equation}
 \label{recursion}
{\cal A}^{(n)}_{2m}=\frac{(2m-1)}{(2m-2+n)}{\cal A}^{(n)}_{2m-2}
\end{equation}
where $n$ denotes the dimension of the space $n=1,2,3,..$. This recursion can be applied $m$-times
linking ${\cal A}^{(n)}_{2m}$ with
${\cal A}_0^{(n)}=O_n(1)$ which is the surface of the unit sphere. In this way we obtain

\begin{equation}
 \label{recureval}
{\cal A}^{(n)}_{2m}= \frac{(2m-1)..\times 3\times 1}{(2m-2+n)..\times(n+2)\times n}{\cal A}^{(n)}_{0}
\end{equation}

The products (each containing $m$ factors) in the nominator and the denominator can be written as
\begin{equation}
 \label{products1}
(2m-1)..\times 3\times 1 =\frac{(2m)!}{2^mm!}
\end{equation}
and
\begin{equation}
 \label{products2}
(2m-2+n)..\times(n+2)\times
n =2^m\frac{(\frac{n}{2}-1+m))!}{(\frac{n}{2}-1)!}
\end{equation}

So we obtain for (\ref{recureval})
\begin{equation}
 \label{recurevalend}
 {\cal A}^{(n)}_{2m}=\frac{(2m)!(\frac{n}{2}-1)!}{2^{2m}m!(\frac{n}{2}-1+m)!}O_n(1)
\end{equation}
where $O_n(1)$ denotes the surface area of the unit-sphere embedded into the $n$-dimensional space \cite{abramowitz}

\begin{equation}
 \label{unit-spheren}
 O_n(1)=\frac{2\pi^{\frac{n}{2}}}{(\frac{n}{2}-1)!}=\frac{2\pi^{\frac{n}{2}}}{\Gamma(\frac{n}{2})}
\end{equation}
and hence
\begin{equation}
 \label{recurevalendfinal}
 {\cal A}^{(n)}_{2m}=\frac{2\pi^{\frac{n}{2}}(2m)!}{2^{2m}m!(\frac{n}{2}-1+m)!} =\frac{2\pi^{\frac{n}{2}}(2m)!}{2^{2m}m!\Gamma(\frac{n}{2}+m)}
\end{equation}

So we can perform the unit-sphere surface integral of any sufficiently smooth function $f(\tau)$

\begin{equation}
\label{evenfu}
f(\xi n_1)=\sum_{m=0}^{\infty}\frac{a_m}{m!}\xi^m n_1^m
\end{equation}

\begin{equation}
 \label{anyfu}
g(\xi)=\int_{|{\bf n}|=1}{\rm d}\Omega({\bf n})f(\xi n_1)
\end{equation}
where only its even part $\frac{f(\tau)+f(-\tau)}{2}$ (even powers) contributes. By using (\ref{integm})-(\ref{recurevalendfinal})
we obtain for (\ref{anyfu}) the series

\begin{equation}
\label{obtainares}
g(\xi)= \sum_{m=0}^{\infty}a_{2m}\frac{{\cal A}^{(n)}_{2m}}{(2m)!}\xi^{2m}
\end{equation}
which is uniquely determined by all even derivatives $a_{2m}$ of $f(\tau)$ at $\tau=0$.
\newline\newline \noindent {\bf Explicit evaluation of (\ref{spheric})}
\newline\noindent
With (\ref{obtainares}) we can evaluate the function $G_n(i\xi)$ which is defined by surface integral
(\ref{spheric}). The integrand (\ref{evenfu}) is $f(\xi n_1)=\frac{1}{(2\pi)^n}\cos{\xi n_1}$ (or alternatively
$f(\xi n_1)=\frac{1}{(2\pi)^n}e^{i\xi n_1}$ since any odd function such as $\sin{\xi n_1}$ yields a zero contribution). Then we obtain
for $G_n(i\xi)$ with (\ref{obtainares}) the following series

\begin{equation}
 \label{sriesG}
G_n(i\xi)=\frac{2}{(4\pi)^{\frac{n}{2}}} \sum_{m=0}^{\infty}(-1)^m \frac{1}{m!\Gamma(m+\frac{n}{2})}
\frac{\xi^{2m}}{2^{2m}}
\end{equation}

Taking into account the generalized definition of the Bessel function of the first kind (\cite{abramowitz} p. 360, 9.1.10.) for any (integer and non-integer) $\nu \in \R$

\begin{equation}
 \label{bessel}
J_{\nu}(\xi) = \left(\frac{\xi}{2}\right)^{\nu}\sum_{m=0}^{\infty}\frac{1}{m!\Gamma(m+\nu+1)}\left(\frac{-\xi^2}{4}\right)^m
\end{equation}

So we obtain for (\ref{spheric}) the closed form expression

\begin{equation}
 \label{Gnclosed}
G_n(i\xi)= \frac{1}{(2\pi)^{\frac{n}{2}}
\xi^{\frac{n}{2}-1}}J_{\frac{n}{2}-1}(\xi)
\end{equation}
which holds for any dimensions $n\in \N$ and is even valid for positive non-integer dimensions since
(\ref{recurevalend}) with (\ref{unit-spheren}) assume the same form when they are deduced for non-integer $n>0$.
Let us now look especially on the physically important cases of dimensions $n=1,2,3$.\newline\newline
\noindent {\bf (i) $\bf n=1$}\newline \noindent
We have with
\begin{equation}
 \label{faculties} 2^{2m}m!(m-\frac{1}{2})!=(2m)(2m-2)..2\times (2m-1)(2m-3)..3 \Gamma(\frac{1}{2})=(2m)!\sqrt{\pi}
\end{equation}
and hence so
\begin{equation}
\label{besselnhalf}
J_{-\frac{1}{2}}(\xi)=\sqrt{\frac{2}{\xi}\pi}\sum_{m=0}^{\infty}(-1)^m\frac{\xi^{2m}}{(2m)!}=
\sqrt{\frac{2}{\pi\xi}}\cos{\xi}
\end{equation}
so that
\begin{equation}
 \label{G1n1}
G_{n=1}(i\xi)=\frac{1}{\pi}\cos{\xi}
\end{equation}
which we reverify to be correct by its definition $G_{n=1}(\xi)=\frac{1}{2\pi}\left(e^{i\xi}+e^{-i\xi}\right)$.
\newline\newline
\noindent {\bf (ii) $\bf n=2$}\newline
\noindent Then we have
\begin{equation}
 \label{besselnindexnull}
J_0(\xi)=\sum_{m=0}^{\infty}(-1)^m\frac{\xi^{2m}}{2^{2m}m!m!}
\end{equation}
and
\begin{equation}
 \label{G1n2}
G_{n=2}(i\xi)=\frac{1}{(2\pi)}J_0(\xi)
\end{equation}
which we can again reverify directly by the definition of $G_2$
\begin{equation}
 \label{deG2}G_2(i\xi)=\frac{1}{(2\pi)^2}\int_0^{2\pi}e^{i\xi\cos{\varphi}}{\rm d}\varphi
\end{equation}
by taking into account the definition of $J_0$.
\newline\newline
\noindent {\bf (iii) $\bf n=3$}
\newline\noindent
Finally we have
\begin{equation}
 \label{facultieseval}
2^{2m}m! (m+\frac{1}{2})! =(2m+1)(2m-1)\times .. 3 \left(\frac{1}{2}\right)!\times 2m(2m-2)\times..2=(2m+1)!\left(\frac{1}{2}\right)! = (2m+1)!\frac{\sqrt{\pi}}{2}
\end{equation}
as $\left(\frac{1}{2}\right)!=\frac{1}{2}\Gamma(\frac{1}{2})= \frac{\sqrt{\pi}}{2}$
and hence
\begin{equation}
 \label{besselfunhalb}
J_{\frac{1}{2}}(\xi)=\sqrt{\frac{2\xi}{\pi}}\sum_{m=0}^{\infty}(-1)^m\frac{\xi^{2m}}{(2m+1)!}=\sqrt{\frac{2}{\pi\xi}}\sin{\xi}
\end{equation}
so that we obtain
\begin{equation}
G_{n=3}(i\xi)= \frac{1}{2\pi^2}\frac{\sin{\xi}}{\xi}
\end{equation}
which again is directly verifiable by the definition
\begin{equation}
\label{G3def}
G_3(i\xi)=\frac{1}{(2\pi)^3}\int_0^{2\pi}\int_0^{\pi}e^{i\xi\cos{\theta}}\sin{\theta}{\rm d}\theta{\rm d}\varphi=
\frac{1}{(2\pi)^3}4\pi\int_0^1\cos{(\xi u)}{\rm d}u =\frac{1}{2\pi^2}\frac{\sin{\xi}}{\xi} .
\end{equation}

\end{document}